\documentclass[10pt,aps,prx,twocolumn,superscriptaddress]{revtex4-2}
\usepackage{amsmath,amssymb}
\usepackage{bm,mathtools}
\usepackage{graphicx}
\usepackage{float}
\usepackage{txfonts}
\usepackage{braket}
\usepackage{algorithm}
\restylefloat{algorithm}
\usepackage{algpseudocode}

\usepackage{tikz}
\usetikzlibrary{positioning,arrows.meta,fit,backgrounds}
\usetikzlibrary{calc}

\usepackage{siunitx}
\usepackage{xcolor}
\usepackage{booktabs}
\usepackage{orcidlink}
\usepackage[section]{placeins}
\usepackage{multirow}

\usepackage{xurl}
\usepackage{hyperref}

\algdef{SE}[SUBALG]{Indent}{EndIndent}{}{\algorithmicend\ }%
\algtext*{Indent}
\algtext*{EndIndent}

\begin{document}
\title{Quantum optimization beyond QUBO for industrial logistics and scheduling}

\author{Juan F. R. Hern\'andez$^{\orcidlink{0009-0007-6933-186X}}$}
\email{juanfrh7@icloud.com}
\affiliation{Kipu Quantum GmbH, Greifswalderstrasse 212, 10405 Berlin, Germany}

\author{Pavle Nika\v{c}evi\'c$^{\orcidlink{0000-0002-8832-5541}}$}
\email{pavle.nikacevic@kipu-quantum.com}
\affiliation{Kipu Quantum GmbH, Greifswalderstrasse 212, 10405 Berlin, Germany}

\author{Enrique Solano$^{\orcidlink{0000-0002-8602-1181}}$}
\affiliation{Kipu Quantum GmbH, Greifswalderstrasse 212, 10405 Berlin, Germany}

\author{Chinonso Onah}
\affiliation{Volkswagen AG, Berliner Ring 2, 38440 Wolfsburg, Germany}

\author{Agneev Guin}
\affiliation{Volkswagen AG, Berliner Ring 2, 38440 Wolfsburg, Germany}

\author{Arne-Christian Voigt}
\affiliation{Volkswagen AG, Berliner Ring 2, 38440 Wolfsburg, Germany}

\author{Archismita Dalal$^{\orcidlink{0000-0003-0638-8328}}$}
\affiliation{Kipu Quantum GmbH, Greifswalderstrasse 212, 10405 Berlin, Germany}

\date{\today}

\begin{abstract}
The increasing complexity of industrial scheduling and transport routing problems motivates the study of alternative optimization formulations and computational paradigms. In this work, we study how higher-order unconstrained binary optimization (HUBO) formulations of such problems map onto quantum optimization workflows in both noisy and fault-tolerant regimes. We consider three representative logistics and manufacturing use cases and formulate each as a HUBO problem. This captures process intricacies, such as highly correlated assembly-line scheduling rules, which are difficult to express faithfully with the standard quadratic (QUBO) form, while at the same time reducing the number of binary variables required in the quantum mapping, thus lowering qubit demand. We compare the HUBO formulations with corresponding QUBO encodings, highlighting a key trade-off: while HUBO reduces qubit requirements through compact binary encoding, it introduces higher-order interaction terms that increase circuit depth, limiting feasibility on current quantum hardware. The proposed formulations are validated using classical solvers across several problem instances and benchmark small routing problem instances using bias-field digitized counterdiabatic quantum optimization in classical simulation. We complement these results with a resource and scalability analysis, focusing on the capacitated vehicle routing problem as a representative large-scale industrial use case. Our analysis indicates that while HUBO formulations offer advantages in qubit scaling compared to QUBO encodings, their practical implementation is constrained by gate fidelity, coherence, and circuit depth, making hybrid quantum--classical workflows and early fault-tolerant quantum hardware the most plausible settings for their practical use.
\end{abstract}

\maketitle

\section{Introduction}
Quantum computing for industry-relevant combinatorial optimization has so far been studied predominantly through the lens of quadratic unconstrained binary optimization~(QUBO), which maps naturally to Ising Hamiltonians and underpins most quantum-annealing and gate-based optimization workflows~\cite{lucas2014ising, farhi2014qaoa, hauke2020perspectives}. 
This focus is primarily hardware-dependent, as commercially available quantum annealers natively support only quadratic couplings, so higher-order unconstrained binary optimization~(HUBO) must be reduced to QUBO using auxiliary variables, often inflating the problem size~\cite{boros2002pseudo, chancellor2017qubo, rosenberg1975reduction}. 
Likewise, while digital algorithms such as quantum approximate optimization algorithm~(QAOA) can in principle represent higher-order interactions, near-term implementations on noisy hardware typically become depth-limited as the interaction order and problem size grow \cite{hadfield2019qaoa, cerezo2021variational}. 
A recent work~\cite{chandarana2025runtime} empirically demonstrated runtime advantages of digitized counterdiabatic quantum optimization~(DCQO) over state-of-the-art classical optimizers across several HUBO instances, including the challenging LABS benchmark  \cite{koch2025quantum}. 

In this work, rather than focusing on algorithmic performance, we investigate the formulation and implementability of higher-order optimization problems arising in industry-relevant settings, with particular emphasis on the tradeoff between representational compactness and implementation cost.
In particular, we reformulate relevant optimization use cases from industrial logistics and manufacturing workflows as HUBO problems, and assess the potential benefits and the practical aspects of current and emerging quantum hardware. 

Quantum computing has been explored for NP-hard scheduling and routing tasks across both quantum annealing and gate-based paradigms. Early feasibility for job-shop scheduling via QUBO encoding was demonstrated using D-Wave quantum annealers~\cite{Venturelli2015}, while systematic evaluations report that although small and constrained instances can be solved and sometimes match classical heuristics, embedding overhead and limited connectivity quickly saturate current devices \cite{Carugno2022}. 
To mitigate hardware constraints, hybrid quantum--classical pipelines and problem decomposition are increasingly used. For example, a multi-objective flexible job-shop scheduler leveraging annealing improved solution quality over a classical baseline benchmark \cite{Schworm2024}. 
On gate-based hardware, tailored variational approaches have been applied to scheduling, with case studies showing advantages over off-the-shelf QAOA on modest-size instances executed on real processors \cite{Amaro2022}. 
In routing, surveys document rapid growth of quantum methods, with traveling salesman problem (TSP) \cite{tsp} dominating and vehicle routing problem (VRP) gaining traction \cite{Osaba2022}. Recent work solved a rich, real-world package-delivery VRP using D-Wave's hybrid solver \cite{Osaba2024}, while QAOA formulations for heterogeneous VRP provide insights into parameterization and scaling on small problem sizes \cite{Fitzek2024}. 
Collectively, these studies mark progress toward practical quantum optimization for scheduling and routing, yet key gaps remain: current demonstrations are limited in scale, solution quality is often comparable, rather than superior, to strong classical heuristics, and performance is constrained by noise, shallow circuits, and QUBO embedding overheads. 
These limitations motivate the exploration of alternative problem formulations that better balance expressivity and hardware requirements.

In real-world routing and scheduling applications, the natural mathematical structure of the problem often leads to HUBO formulations rather than purely quadratic ones. Higher-order polynomial terms may emerge from rich variable encodings, nonlinear objectives, and the need to faithfully impose intricate operational constraints. Such HUBO models are especially relevant in industrial settings, where oversimplifying to QUBO can obscure key dependencies and lead to less faithful representations of the underlying operational constraints.
To illustrate the range and utility of HUBO approaches, we present three representative use cases drawn from different industry-relevant logistics and manufacturing scenarios.
For the two transportation-routing cases, we consider the quantum-enhanced shared transportation (QUEST) problem~\cite{Onah2025QUEST,Onah2026QUEST2}, and the capacitated vehicle routing problem (CVRP) \cite{dantzig1959truck,toth2014vehicle}. 
Our HUBO formulations reduce qubit scaling from linear to logarithmic, improving representational compactness for a fixed qubit budget.
This reduction in qubit requirements, however, comes at the cost of increased two-qubit gate requirement, which becomes a key limiting factor on current hardware.
While near-term noisy devices still restrict us to small instances, these results suggest that HUBO formulations may become more attractive than QUBO in qubit-limited early fault-tolerant settings, provided that circuit-depth overheads can be controlled~\cite{Romero2025BFDCQO}. 
For the production scheduling case, where realistic instance sizes exceed current hardware and simulation limits, we therefore complement the HUBO modeling with a quantum--classical hybrid workflow that can leverage near-term quantum solvers on localized subproblems while preserving the higher-order structure of the full industrial objective.

Our goal is therefore not to demonstrate quantum advantage on industrial-scale instances, but to clarify when higher-order formulations offer a meaningful representational advantage and how that advantage is constrained by quantum resource requirements.
The paper is organized as follows. In \S\ref{sec:quantum_opt}, we introduce quantum optimization methods, with a focus on the bias-field DCQO~(BF-DCQO) algorithm and its circuit-fidelity and runtime estimates. In \S\ref{sec:routing}, we first present the two routing use cases (QUEST and CVRP), including their HUBO formulations and validation using classical solvers such as simulated annealing (SA), together with BF-DCQO solutions for small instances.
Furthermore in \S\ref{sec:scalability}, we analyze the scalability of the CVRP formulations in terms of qubit and circuit depth, and assess the practical feasibility of BF-DCQO algorithm for this use case in the early fault-tolerant regime.
In \S\ref{sec:scheduling}, we introduce the automotive scheduling use case, describe its HUBO formulation, and evaluate it using classical solvers and a hybrid rolling-horizon workflow, which is based on classical subproblem decomposition and optimization.  
Finally, we conclude with an outlook in \S\ref{sec:conclusion}.

\section{Quantum optimization methods}
\label{sec:quantum_opt}
Classical optimization methods, such as gradient-based algorithms, mixed-integer programming, and simulated annealing, rely on deterministic or stochastic exploration of the search space \cite{nocedal2006numerical, conforti2014integer, kirkpatrick1983optimization}. These methods face exponential growth in computational cost as problem size increases, making large combinatorial instances intractable. Quantum optimization algorithms address this challenge by exploiting quantum phenomena to explore multiple configurations simultaneously, offering the potential for faster convergence and improved solution quality compared to classical approaches.

Techniques such as Grover search \cite{10.1145/237814.237866}, Adiabatic Quantum Computing (AQC) \cite{farhi2000quantumcomputationadiabaticevolution}, and quantum annealing \cite{PhysRevE.58.5355} are promising candidates to achieve quantum advantage over classical methods. These algorithms were originally designed for fault-tolerant quantum computers, where noise and errors are effectively suppressed to ensure reliable computations. To address the limitations of current noisy intermediate-scale quantum (NISQ) devices, hybrid algorithms such as the Quantum Approximate Optimization Algorithm (QAOA) \cite{farhi2014qaoa} have been developed, combining classical optimization loops with shallow quantum circuits inspired by AQC principles \cite{lloyd2018quantumapproximateoptimizationcomputationally}. However, QAOA requires classical optimization of circuit parameters, often leading to barren plateaus. Moreover, it usually requires long circuit depths to achieve high-quality solutions \cite{SciRep2024_QAOA_Depth_Noise, PhysRevResearch2025_QAOA_Limitations_Generic}, which often exceed the coherence time of present-day quantum hardware, leading to noisy results. 

A promising alternative is offered by counterdiabatic (CD) approaches \cite{hegade2021shortcuts, hegade2022digitized}. By adding auxiliary CD terms into the adiabatic Hamiltonian, the non-adiabatic transitions are suppressed, enabling faster evolution with fewer Trotter steps and hence reduced circuit depth. In particular, we implement bias-field Digitized Counterdiabatic Quantum Optimization (BF-DCQO) \cite{Cadavid_2025}, designed to solve combinatorial optimization problems on current digital quantum hardware. 

\subsection{Bias-field DCQO}

This protocol extends standard DCQO \cite{dcqo} by introducing an iterative feedback mechanism based on longitudinal bias fields. After each evolution, the system's measurement outcomes are used to update a bias field that progressively guides the dynamics toward lower-energy configurations until convergence. The total Hamiltonian governing the evolution is
\begin{equation}
    H(\lambda) = [1 - \lambda(t)] \tilde{H}_{\text{initial}} + \lambda(t) H_{\text{final}} + H_{\text{cd}},
\end{equation}
where $\lambda(t)$ is a time-dependent control parameter defining the adiabatic path, such that $\lambda(0)=0$ and $\lambda(T)=1$, with $T$ being the total evolution time. The scheduling function is chosen as $\lambda (t) = \sin^2 \left[\frac \pi 2 \sin^2\left(\frac {\pi t}{2T}\right)\right]$, so that the counterdiabatic terms vanish at the beginning and at the end of the evolution.

Instead of initializing the system with the standard transverse-field Hamiltonian $H_{\text{initial}} = \sum_{i=1}^{N} h_i^x \sigma_i^x$, whose ground state $|+\rangle^{\otimes N}$ can be easily prepared, BF-DCQO introduces a longitudinal bias field derived from previous iterations. The modified initial Hamiltonian is defined as
\begin{equation}
    \tilde{H}_{\text{initial}} = \sum_{i=1}^{N} \left( h_i^x \sigma_i^x - \tilde{h}_i^b \sigma_i^z \right),
\end{equation}
where $\tilde{h}_i^b = \langle \sigma_i^z \rangle$ is obtained by measuring the qubits in the computational basis after each iteration. This feedback mechanism effectively biases the system toward low-energy configurations found in earlier runs, thereby accelerating convergence to the global minimum.

The final Hamiltonian encodes the combinatorial optimization problem in the form of a HUBO problem,
\begin{equation}
    H_{\text{final}} = \sum_{i=1}^{N} h_i \sigma_i^z + \sum_{i<j}^{N} J_{ij} \sigma_i^z \sigma_j^z + \sum_{i<j<k}^{N} K_{ijk} \sigma_i^z \sigma_j^z \sigma_k^z + \dots,
\end{equation}
where the coefficients $h_i$, $J_{ij}$, and $K_{ijk}$ define the problem-specific couplings. The ground state of $H_{\text{final}}$ represents the optimal solution of the target optimization instance.

To suppress non-adiabatic transitions that may arise from finite-time evolution, a counterdiabatic correction is incorporated into the Hamiltonian,
\begin{equation}
    H_{\text{cd}} = \dot{\lambda}(t) A_{\lambda},
\end{equation}
where $A_{\lambda}$ denotes the adiabatic gauge potential \cite{KOLODRUBETZ20171}. Obtaining an exact expression for $A_{\lambda}$ is in general intractable for many-body systems due to its exponential complexity. To address this, BF-DCQO employs an approximate construction using the nested commutator method, which expresses $A_{\lambda}$ as a truncated series expansion whose coefficients are determined variationally \cite{doi:10.1073/pnas.1619826114, PhysRevLett.123.090602, PhysRevA.103.012220, PhysRevB.106.155153, PhysRevX.14.011032}. This approach provides a first-order approximation that captures the leading counterdiabatic corrections with modest resource requirements.

In practice, each iteration of BF-DCQO proceeds as follows. The system is initialized in the biased ground state of $\tilde{H}_{\text{initial}}$, which can be prepared using single-qubit $R_y(\theta_i)$ rotations, where $\theta_i=2\arctan\left(\frac{h_i^x}{\tilde{h}_i^b-\sqrt{(\tilde{h}_i^b)^2+(h_i^x)^2}}\right)$ is determined by the bias parameters $\tilde{h}_i^b$ and the initial transverse field Hamiltonian $h_i^x=-1$. The digitized counterdiabatic evolution under $H(\lambda)$ is then implemented using a finite number of Trotter steps. After the evolution, all qubits are measured in the computational basis to estimate the expectation values $\langle \sigma_i^z \rangle$, which are subsequently used to update the bias fields for the next iteration. This feedback loop gradually steers the quantum state distribution toward the ground state of $H_{\text{final}}$ while maintaining a shallow circuit depth. In our simulations, we use a two-step discretization of the evolution schedule because $\dot{\lambda}(t)$ vanishes at the endpoints, reducing the problem to a single effective counterdiabatic step.

The BF-DCQO protocol has demonstrated a polynomial scaling advantage in the ground-state success probability compared to both standard DCQO and finite-time adiabatic quantum optimization \cite{bfdcqo}. It achieves faster convergence and higher-quality solutions using significantly fewer quantum gates. Moreover, since it is a fully quantum algorithm that does not rely on classical optimization loops, BF-DCQO avoids the trainability issues commonly observed in variational quantum algorithms. Its robustness has been validated through experiments on both trapped-ion (IonQ Forte) and superconducting (IBM Brisbane) quantum processors, showing up to two orders of magnitude improvement in success probability and approximation ratio over QAOA for dense HUBO instances \cite{chandarana2025runtime, Romero2025BFDCQO}.

\subsection{Circuit fidelity and runtime estimates for BF-DCQO algorithm}
\label{sec:resource}
Now we discuss the gate fidelity and gate duration requirements for the BF-DCQO algorithm, and how they impact the algorithm's runtime.

For a $N$-qubit DCQO circuit with $n_\text{2q}$ two-qubit gates and an average two-qubit gate fidelity of $f_\text{2q}$, the overall circuit fidelity can be estimated as
\begin{equation}
\label{eq:dcqo_fid}
    f_\text{DCQO} \approx f_\text{2q}^{n_\text{2q}}(f_\text{SPAM}f_\text{1q})^N\;
    \approx (1-e_\mathrm{2q})^{n_\text{2q}},
\end{equation}
where the average two-qubit gate error~$e_\mathrm{2q}$ is approximated as $1-f_\text{2q}$.
We can assume that the total SPAM (state preparation and measurement) errors and one-qubit gate errors are negligible as compared to the contribution from two-qubit gates.
This expression determines the maximum amount of tolerable two-qubit gate errors for a given DCQO circuit. 

This also highlights the rapid decay of circuit fidelity with increasing two-qubit gate count, indicating that low error rates are essential for scaling to larger problem instances.

We estimate the overall execution cost of BF-DCQO by separating classical and quantum contributions,
\begin{equation}\label{eq:2q_fidelity}
T_{\mathrm{BF\text{-}DCQO}} = T_{\mathrm{CPU}} + T_{\mathrm{QPU}} .
\end{equation}
The classical subroutine typically involves bit-flip corrections, and the corresponding runtime is approximated as
\begin{equation}
T_{\mathrm{CPU}}
=
n_{\mathrm{CVaR}} n_{\mathrm{iter}} n_{\mathrm{sweep}}
T_\mathrm{sweep}\ \mathrm{s},
\end{equation}
where CVaR dictates what percentage of measurements are used to calculate the bias values. $T_\mathrm{sweep}$ depends on the problem as it involves calculating the cost function. The
quantum execution time is calculated as
\begin{equation}
T_{\mathrm{QPU}}
=
n_{\mathrm{iter}} n_{\mathrm{shots}}T_\mathrm{shot}\ \mathrm{s}.
\end{equation}
We omit circuit compilation and transpilation overhead in
this estimate because we assume all-to-all connected quantum architecture. We estimate $T_\mathrm{shot}$ to be:
\begin{equation}
    T_\mathrm{shot} \approx t_\mathrm{2q}d + t_\mathrm{reset} \; \ge t_\mathrm{2q}(2n_\mathrm{2q}/N) + t_\mathrm{reset},
\label{eq:shot}
\end{equation}
which strongly depends on the quantum hardware modality.
The lower-bound depth estimate assumes maximal parallelism on all-to-all connectivity.

Under a global depolarizing noise model, the measured Pauli expectation, i.e.\ the bias magnitudes in BF-DCQO, is attenuated relative to the ideal value as
$\langle Z\rangle_{\mathrm{noisy}} = \eta\,\langle Z\rangle_{\mathrm{ideal}}$,
with $\eta\in(0,1]$ capturing the circuit-level fidelity loss, i.e.\ $f_\mathrm{DCQO}$ in this context. 
Estimating $\langle Z\rangle_{\mathrm{ideal}}$ by rescaling amplifies statistical fluctuations by $1/\eta$, so the required number of shots to achieve $\epsilon$-accuracy, with failure probability $\delta$, is given by~\cite{HKP20}
\begin{equation}\label{eq:n_shots}
n_{\mathrm{shots}} \sim \frac{1}{f_\mathrm{DCQO}^{2}}\;\frac{1}{\epsilon^{2}}\;\log\!\left(\frac{N}{\delta}\right),
\end{equation}
where $N$ is also the number of simultaneously estimated $\langle Z\rangle$ observables, i.e. bias values, and the factor $1/f_\mathrm{DCQO}^{2}$ is the sampling overhead induced by noise.

A key limitation of end-to-end quantum optimization arises from the interplay between circuit fidelity and execution time. In the NISQ regime, the number of required shots scales inversely with the square of the circuit fidelity, which itself decays exponentially with circuit depth, leading to a rapid growth in sampling cost for larger instances. In contrast, fault-tolerant implementations suppress this effect through high logical fidelity, but incur significantly increased per-shot execution times due to the overhead of error correction. 

To obtain practically relevant runtimes, we can adopt a sequential approach, in which the quantum routine is used as a refinement step within a predominantly classical optimization pipeline~\cite{2025arXiv251005851C}. 
In this setting, we can assume a fixed shot budget and can restrict the circuit depth through various techniques including truncation of higher-order or less-relevant terms and partial encoding of variables. This allows the runtime of the quantum subroutine to remain controlled, while classical preprocessing and postprocessing handle global exploration and refinement of candidate solutions.

\section{Routing problems}
\label{sec:routing}
Routing problems constitute a fundamental class of combinatorial optimization tasks with broad relevance in logistics, transportation, and industrial operations. In general, these problems seek an optimal way to connect, order, or assign locations while minimizing a cost function and satisfying operational constraints. Among them, the Traveling Salesman Problem (TSP) is the canonical benchmark: given a set of cities and pairwise distances, the objective is to find the minimum-cost tour that visits each city exactly once and returns to the starting point \cite{lawler1985traveling,applegate2006traveling}.

We consider two routing problems that can be viewed as structured extensions of the TSP: the first is the Quantum-Enhanced Shared Transportation (QUEST) problem~\cite{Onah2025QUEST,Onah2026QUEST2}, and the second is the Capacitated Vehicle Routing Problem (CVRP) \cite{dantzig1959truck,toth2014vehicle}. These problems provide natural test cases for assessing how HUBO formulations can trade increased interaction order with variable reductions relative to standard QUBO encodings.

\subsection{Use case I: QUEST}
\label{sec:quest}
The QUEST framework is an example of a quantum framework for a vehicle matching optimization problem. In this use case, each vehicle categorized as a windbreaker~(breaker) is assigned to a windsurfer~(surfer), where the surfer~$s$ is supposed to drive behind the breaker~$b$ on a shared road segment and benefit from a reduced aerodynamic cost.
\begin{figure*}
    \centering
    \includegraphics[width=\textwidth]{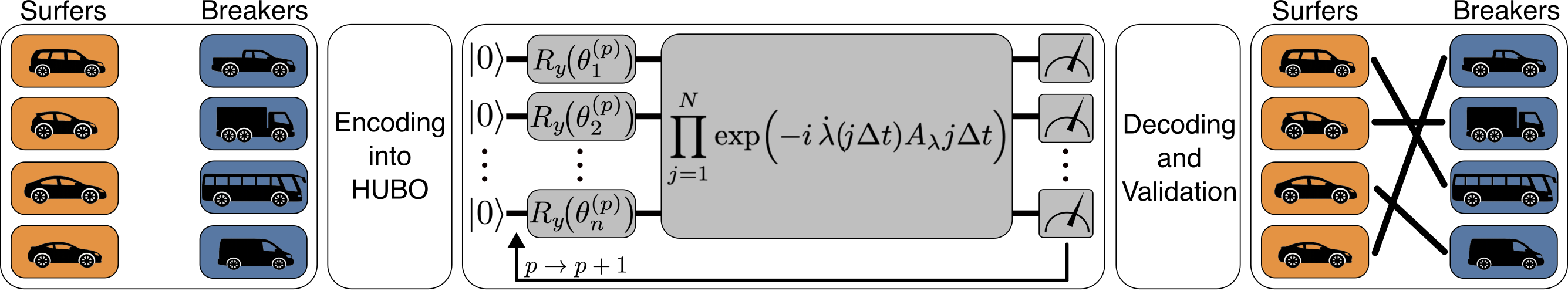}
    \caption{High-level QUEST workflow from the set of breakers and surfers, through HUBO encoding and iterative BF-DCQO evolution, to decoding and validation of the final breaker--surfer matching. In the central panel, the quantum circuit represents the $p$-th BF-DCQO iteration: single-qubit $R_y$ rotations prepare the biased initial state, the digitized counterdiabatic evolution implements the problem Hamiltonian, and measurement outcomes are used to update the bias field for the next iteration.}
    \label{fig:bf-dcqo-flow}
\end{figure*}

\subsubsection{Problem formulation}
The goal of QUEST is to construct the bijective map from the sets of breakers to the set of surfers such that total cost $\omega_{s,b}$ is minimized. The cost should take into account the aerodynamic drag, but also ensure that the paired breaker and surfer match in their preferred driving speeds and arrival times. The cost is thus defined as
\begin{equation}
    \omega_{s,b} := c_sv_b^2[1-f(c_b-c_s)]+\lambda_1\Delta_{s,b}^t+\lambda_2\Delta_{s,b}^v,
\end{equation}
where $c_i \in \{1,2,3,4,5\}$  represents the size class of vehicle $i$, $v_i$ is its preferred driving speed, $\Delta_{s,b}^t$ and $\Delta_{s,b}^v$ are the differences in arrival times and preferred speeds between $s$ and $b$, respectively, with $\lambda_1$ and $\lambda_2$ being the respective penalty constants (only those differences larger than surfer-specific thresholds are penalized), and $f(d):=\frac{d+4}{24}$ where $d := c_b  -c_s$ and thus $d \in \{-4,...,4\}$.

In this work, we consider the proof-of-concept formulation for a single road segment only. A generalization of our formulation to multiple road segments is straightforward and should scale favorably using quantum optimization methods. On the other hand, such general formulation involves complex combinatorial interactions which make classical optimization approaches infeasible for large-scale systems, highlighting transportation as a practical and relevant domain for near-term quantum optimization research. 

While the original QUEST formulation described the problem as a QUBO \cite{Onah2025QUEST}, we reformulate it as a HUBO task, offering a more compact and expressive representation. Although standard QUBO formulations are widely used, they suffer from scalability issues due to their reliance on one-hot encodings, which require $\mathcal{O}(N^2)$ binary variables where $N:=B=S$ is the total number of breakers (or, equivalently, surfers) and dense quadratic couplings to enforce constraints. This increases hardware demands and often results in energy landscapes that are difficult to optimize efficiently. 

In contrast, HUBO formulations that use binary number encodings reduce the number of binary variables to $\mathcal{O}(N\log N)$, but introduce higher-order terms. This trade-off, as discussed in the space-efficient encoding framework of Glos et al.~\cite{Glos2020SpaceEfficient}, is particularly well-suited for problems like QUEST. In the HUBO formulation, the surfer assigned to breaker $b$ is represented by a binary-encoded variable $x_b=(x_{b,0},\dots,x_{b,K-1})$, where $x_{b,k}\in\{0,1\}$ and $K=\lceil \log_2 S\rceil$. Collecting all binary variables into $\mathbf{x}=\{x_{b,k}\}$, the total QUEST cost function can be written as
\begin{equation}
H(\mathbf{x})= H_{\mathrm{obj}}(\mathbf{x})
+\lambda_{\mathrm{valid}} H_{\mathrm{valid}}(\mathbf{x})\nonumber+\lambda_{\mathrm{unique}}H_{\mathrm{unique}}(\mathbf{x}),
\end{equation}
where $H_{\mathrm{obj}}$ encodes the matching cost, while $H_{\mathrm{valid}}$ and $H_{\mathrm{unique}}$ penalize invalid surfer labels and repeated surfer assignments, respectively; their explicit forms are given in Appendix~\ref{app:quest_HUBO}.

Structurally, QUEST closely resembles TSP, where cities are matched with times, analogously to pairing breakers with surfers. The penalty terms in TSP can be defined identically, making sure that every city is visited. However, unlike in the TSP, where there are pairwise interactions between cities at adjacent times that represent inter-city distances, in QUEST, each breaker's cost depends only on its own surfer assignment. The absence of inter-breaker interactions simplifies the problem's structure, allowing for a HUBO-based formulation that retains the indexing logic of the TSP while remaining computationally more efficient.

\subsubsection{Results: feasibility and validation}
\begin{figure}
    \centering
\includegraphics[width=\columnwidth]{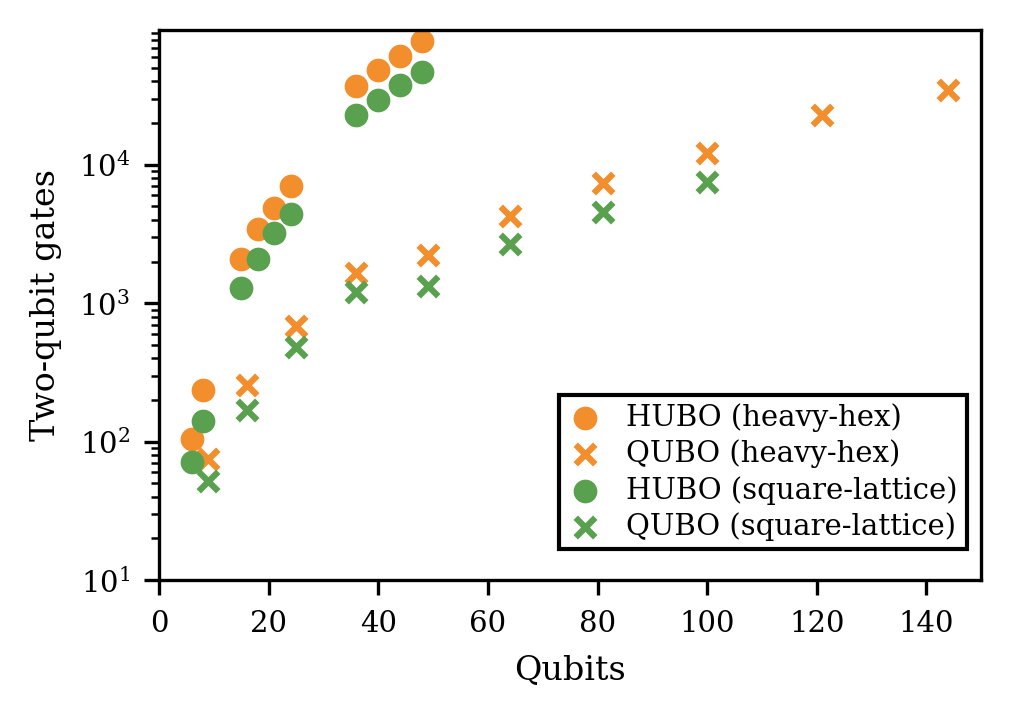}
    \caption{
    Two-qubit gate operations needed to encode QUEST circuits as a function of number of qubits, corresponding to breaker (surfer) counts from 3 to 12. The circuits transpiled to IBM Kingston (heavy-hex) and IBM Miami (square-lattice). In the IBM Miami QUBO case, 11 and 12 breaker instances are not included since they require more qubits than available.}
    \label{fig:quest_scaling}
\end{figure}

We begin by assessing the feasibility of executing the HUBO formulation (Fig.~\ref{fig:bf-dcqo-flow}) on near-term quantum hardware by analyzing the BF-DCQO circuit resources required for our problem instances. 
Throughout this study, we assume a single Trotter step in the BF-DCQO implementation, consistent with empirical evidence from our previous work, where one--step digitized counterdiabatic evolution performed favorably across several QUBO and HUBO benchmarks. As can be seen in Fig.~\ref{fig:quest_scaling}, the HUBO re-formulation offers an advantage in qubit-number scaling relative to a QUBO formulation of the same QUEST instance.

This improvement, however, comes at the cost of an increased entangling-gate count due to higher-order interaction terms. In particular, Fig.~\ref{fig:quest_scaling} highlights the central trade-off between the two encodings. The HUBO formulation compresses the problem into fewer qubits, but the resulting circuits become substantially deeper than their QUBO counterparts because higher-order terms must be decomposed into a larger number of entangling operations. By contrast, the QUBO formulation uses more qubits, yet yields consistently smaller transpiled two-qubit gate counts and is therefore more favorable from the circuit-depth perspective. This distinction is especially relevant for current noisy hardware, where the feasible circuit size is constrained not only by the number of available qubits but also by coherence times and the accumulation of two-qubit gate errors. From Fig.~\ref{fig:quest_scaling}, only the smallest instances of the HUBO formulation appear compatible with near-term execution, whereas instances larger than 7 breakers require two-qubit gates exceeding the limit of current hardware. The figure also showcases a reduction in two-qubit-gate counts with improved connectivity (square-lattice over heavy-hex).
Therefore, for fixed instance sizes, QUBO remains easier to execute on current devices, while the main practical advantage of HUBO lies in its improved qubit scaling, which is more likely to become relevant in the early fault-tolerant regime.

Finally, we validate our HUBO encoding. We compare the approximation ratio for the HUBO encoding using BF-DCQO with the ideal simulator, where we simulated circuits up to 6 breakers and surfers. We define the approximation ratio of a given setup as the ratio between the cost function value at the exact solution (obtained using brute force) and the value obtained by the setup.
Moreover, we use simulated annealing to solve the quadratized HUBO encoding and validate larger instances. For BF-DCQO, we use $3$ bias-field iterations and $10^4$ shots, whereas for simulated annealing, we use $10^6$ sweeps and $10^4$ shots. 
We finally compare these results against the optimal solution. 

\begin{figure}
  \centering
  \includegraphics[width=\linewidth]{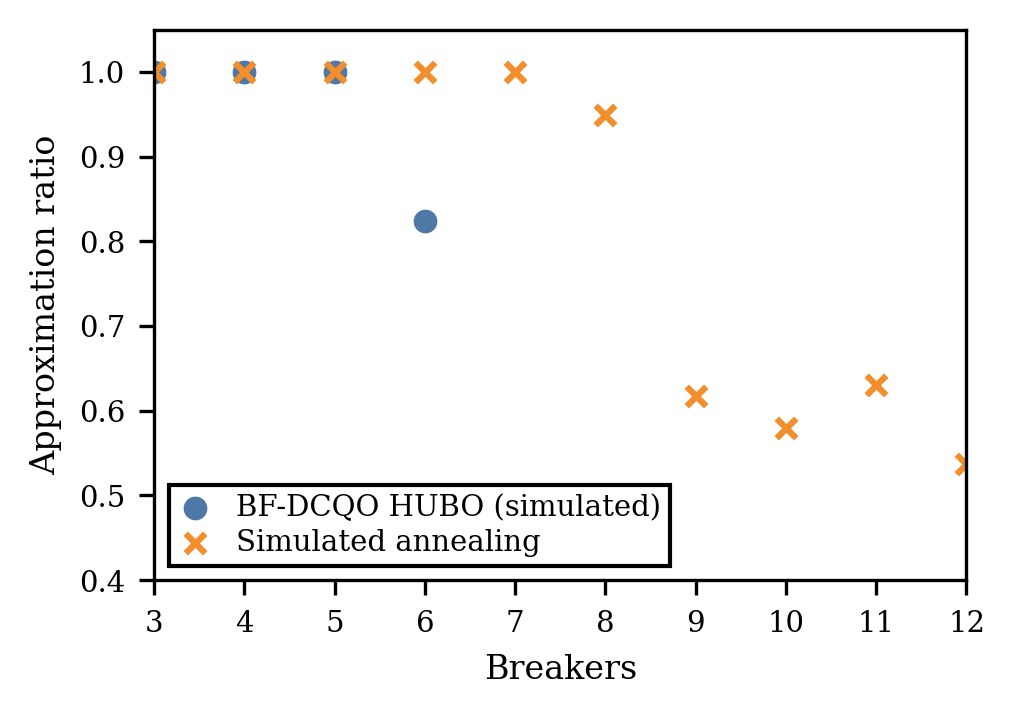}
  \caption{Approximation ratios for simulated annealing quadratized HUBO ($10^4$ shots and $10^6$ sweeps) and BF-DCQO HUBO (statevector simulation, 3 bias-field iterations, $10^4$ shots each). The BF-DCQO HUBO results were obtained with the full Hamiltonian up to 5 breakers, while the instances larger than that used the reduced Hamiltonian.}
  \label{fig:quest_results}
\end{figure}

As shown in Fig.~\ref{fig:quest_results}, both methods recover the optimal solution for the smallest QUEST instances, indicating that the HUBO encoding is correctly capturing the matching problem and that these low-dimensional cases are not yet limited by optimizer performance. As the number of breakers and surfers increases, deviations from the optimum begin to appear. BF-DCQO maintains an approximation ratio of $1$ up to five breakers and surfers, but already drops for the six-breaker instance, which is consistent with the rapid increase in circuit complexity discussed above and suggests that the simulated quantum dynamics becomes more difficult to steer toward the ground state as the higher-order landscape grows. In the case of simulated annealing, it remains optimal up to seven breakers and surfers and then gradually deteriorates, reaching markedly lower approximation ratios for larger instances. This trend indicates that, although the quadratized HUBO can still be handled efficiently by a classical heuristic at intermediate sizes, the optimization landscape becomes increasingly rugged as the problem grows, making it harder for the annealer to consistently identify the global optimum. 
Overall, these results provide a useful benchmark for the proposed QUEST encoding, confirming that the formulation is valid on small instances and illustrating that, as the problem size grows, both the higher-order quantum implementation and the corresponding classical quadratized optimization become increasingly challenging. This supports the view that the compact HUBO formulation introduces a nontrivial optimization landscape, which is precisely the regime of interest for future quantum optimization methods.

\subsection{Use case II: CVRP}
\label{sec:cvrp}

The Vehicle Routing Problem (VRP) \cite{dantzig1959truck} is an important combinatorial optimization problem that aims at determining the most efficient set of routes for a fleet of vehicles that must transport goods from a central depot to a group of customers. It extends the traveling salesman problem, where a single vehicle must visit several locations in the optimal sequence. In this work, we focus on a more practical variant, the Capacitated Vehicle Routing Problem (CVRP), in which all vehicles have a limited capacity for the goods that must be delivered.

\subsubsection{Problem formulation}
We consider a VRP where a single depot is indexed by $0$ and $N$ customers are indexed by $1,\dots, N$, giving a total of $n = N + 1$ nodes. The fleet consists of $M$ vehicles, each capable of visiting up to $L$ customers, where $L$ also represents the maximum number of route positions, or slots, per vehicle. To encode node indices, we use binary encoding, where each node is represented using $K' = \lceil \log_2 n \rceil$ bits, which determines the number of binary variables needed to uniquely identify all nodes. The binary decision variables are denoted by $x_{v,t,k} \in \{0,1\}$, where $v \in \{0,\dots,M-1\}$ indexes the vehicle, $t \in \{0,\dots,L-1\}$ indexes the position within the route, and $k \in \{0,\dots,K'-1\}$ corresponds to the bit position. Collecting all binary variables into $\mathbf{x}=\{x_{v,t,k}\}$, the final CVRP cost function can be written as:
\begin{align}
H(\mathbf{x})= & H_{\mathrm{obj}}(\mathbf{x})
+\lambda_{\mathrm{visit}} H_{\mathrm{visit}}(\mathbf{x}) \nonumber\\
&+\lambda_{\mathrm{mono}}H_{\mathrm{mono}}(\mathbf{x})
+\lambda_{\mathrm{valid}} H_{\mathrm{valid}}(\mathbf{x})
+\lambda_{\mathrm{cap}} H_{\mathrm{cap}}(\mathbf{x}),
\end{align}
where $\lambda_{\mathrm{visit}}$, $\lambda_{\mathrm{mono}}$, $\lambda_{\mathrm{valid}}$, and $\lambda_{\mathrm{cap}}$ are nonnegative penalty coefficients chosen such that any violation of a constraint results in a higher energy value. The individual cost functions are defined in the Appendix \ref{app:CVRP_HUBO}.

This HUBO formulation offers a logarithmic reduction in the number of binary variables compared with the standard QUBO formulation (one-hot encoding). The number of variables scales as $\mathcal{O}(ML\log N)$ for the HUBO case, whereas it scales as $\mathcal{O}(MLN)$ for the QUBO formulation. 

This reduction in variable count, however, comes at the cost of additional higher-order interaction terms. For example, binary encoding can produce unphysical customer indices, which must be excluded through additional penalty terms \ref{eq:valid}. Moreover, binary encoding increases the degree of the interactions in the Hamiltonian. In particular, the terms in Eqs.~\ref{eq:obj}, \ref{eq:cap}, \ref{eq:mono}, and \ref{eq:visit} are quadratic in the selector functions defined in Eq.~\ref{eq:selector}, leading to higher-order interactions among the decision variables. The maximum interaction order is therefore $2K'$. This, in turn, affects the depth of the final quantum circuit. To estimate this cost, we consider the number of required two-qubit gates. An upper bound on the number of distinct $r$-body monomials, for $r \leq 2K'$, can be obtained through a combinatorial counting argument, as detailed in Appendix~\ref{app:polynomial_scaling}.

\subsubsection{Results: feasibility and validation}

\begin{figure}
\centering
\includegraphics[width=\linewidth]{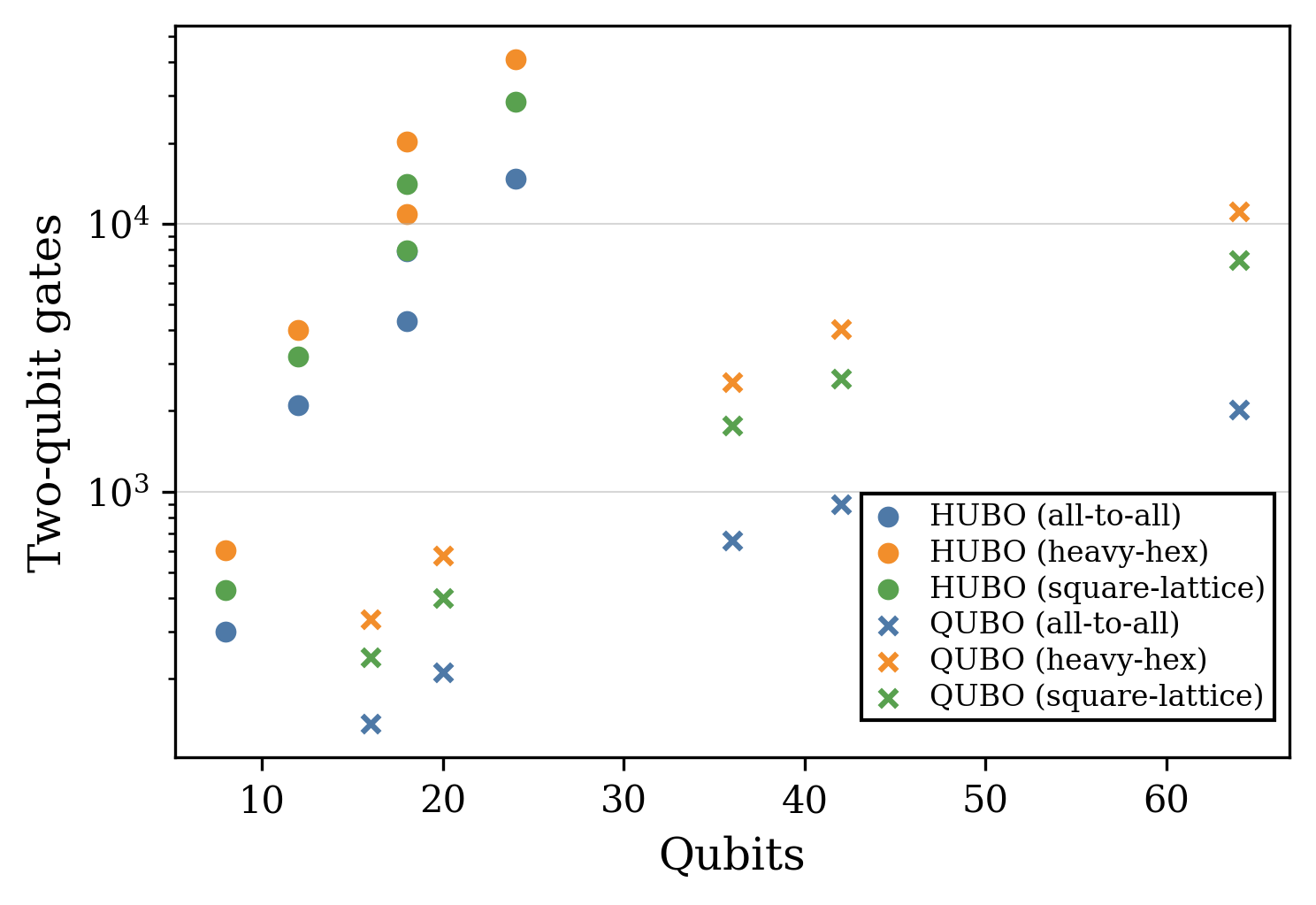}
\caption{Transpiled two-qubit gate counts for a subset of CVRP instances under three connectivity assumptions. The all-to-all results were obtained using the IBM Aer simulator, while the heavy-hex and square-lattice results were obtained by transpiling to IBM Kingston and IBM Miami, respectively.}
\label{fig:cvrp_2q_gate_scaling}
\end{figure}

Figure~\ref{fig:cvrp_2q_gate_scaling} reports the transpiled two-qubit gate counts for a subset of CVRP instances under three different hardware-connectivity assumptions: all-to-all, heavy-hex, and square-lattice connectivity. The results, similar to the QUEST use case, highlight a clear trade-off between the HUBO and QUBO encodings. The HUBO formulation requires fewer qubits than the corresponding QUBO encoding; however, this reduction in qubit count comes at the cost of a larger number of two-qubit gates. Since two-qubit gates are typically the dominant source of noise in near-term quantum devices, the HUBO circuits become more challenging on current hardware, where entangling-gate depth remains a major limitation.

All-to-all connectivity leads to the lowest two-qubit gate counts, since no SWAP operations are required to mediate interactions between distant qubits. This connectivity is characteristic of trapped-ion quantum processors, which currently operate in the $10$--$10^2$ qubit regime. Recent benchmarks have shown experiments containing up to 1083 two-qubit gates, with average two-qubit errors at the $10^{-4}$ level~\cite{hughes2025trappediontwoqubitgates9999}. Under this connectivity model, the HUBO encoding supports instances up to $4$ customers, $2$ vehicles, and capacity $2$, limited by the estimated coherence-time. The QUBO encoding supports instances up to $6$ customers, $2$ vehicles, and capacity $3$, but is instead limited primarily by the required number of physical qubits.

By contrast, heavy-hex and square-lattice connectivities, which are representative of superconducting quantum architectures, require additional SWAP gates to implement non-local interactions. This increases the transpiled circuit depth and, consequently, reduces the size of the circuits that can be realistically executed. The heavy-hex topology introduces a larger two-qubit gate overhead than the square-lattice topology. Nevertheless, superconducting devices currently offer more than $100$ qubits, with up to $5000$ two-qubit gates per circuit~\cite{ibm_nighthawk_2025}. Under these hardware assumptions, the available qubit count allows the simulation of instances up to $5$ customers, $2$ vehicles, and capacity $3$ using the HUBO encoding, and up to $7$ customers, $2$ vehicles, and capacity $4$ using the QUBO encoding.

\begin{figure}
\centering
\includegraphics[width=\linewidth]{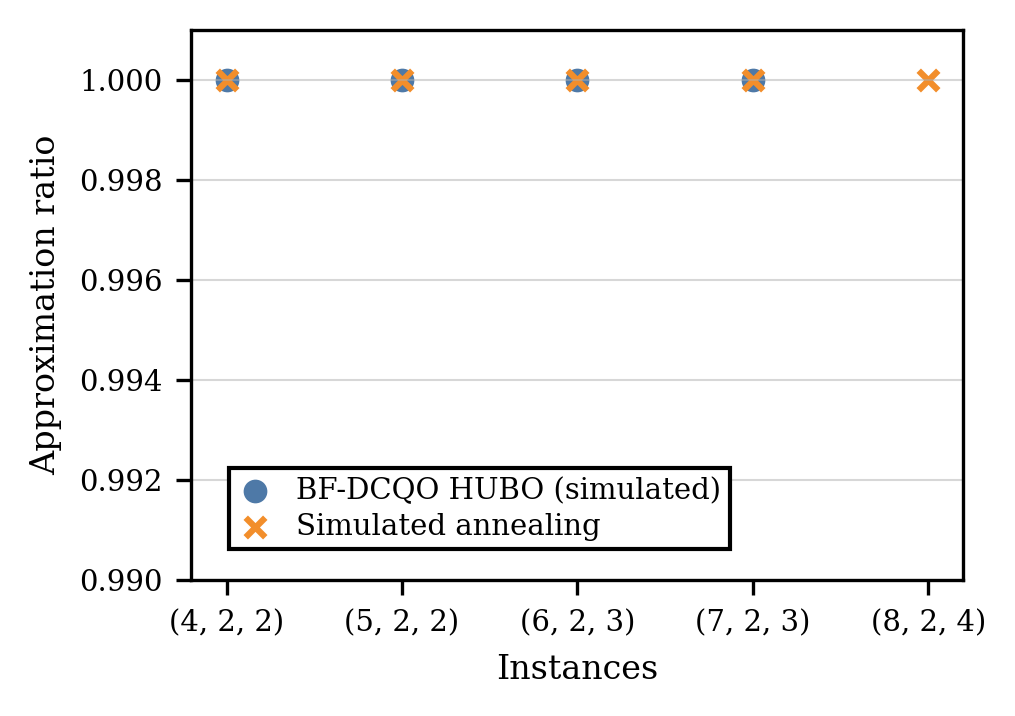}
\caption{Approximation ratios obtained for simulated annealing on the quadratized HUBO formulation using $10^4$ shots and $10^6$ sweeps, and for BF-DCQO on the HUBO formulation using statevector simulation with $10$ bias-field iterations and $10^4$ shots per iteration.}
\label{fig:ar_cvrp}
\end{figure}

Finally, we validate the proposed HUBO encoding by comparing the approximation ratios obtained with two heuristic approaches. First, we solve the HUBO formulation directly using BF-DCQO with a statevector simulator. For BF-DCQO, we use $10$ bias-field iterations and $10^4$ shots per iteration. Second, we solve the quadratized HUBO formulation using simulated annealing, with $10^6$ sweeps and $10^4$ shots. In both cases, the approximation ratio is computed with respect to the optimal solution.
As shown in Figure~\ref{fig:ar_cvrp}, both BF-DCQO applied to the HUBO encoding and simulated annealing applied to the quadratized HUBO encoding recover the optimal solution for all tested instances. These results provide an initial validation of the proposed encoding on small CVRP instances.

\subsection{Scalability analysis}
\label{sec:scalability}

\begin{figure}
\centering
\includegraphics[width=\linewidth]{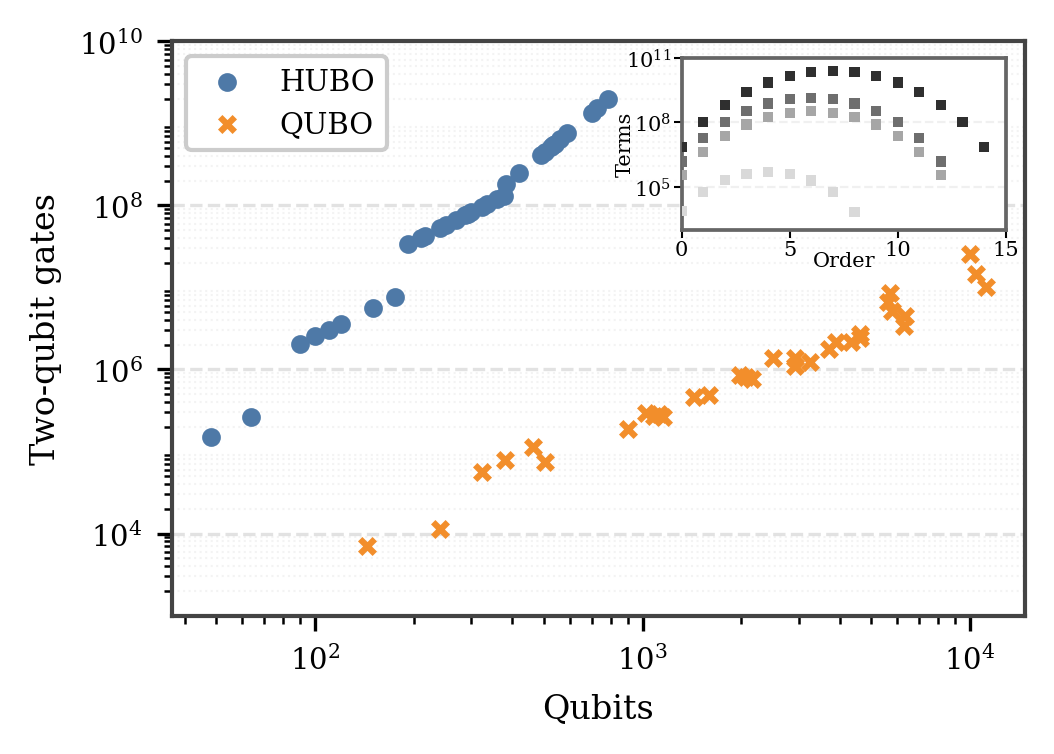}
\caption{
Estimated two-qubit gate counts for a subset of CVRP instances, assuming a single-step DCQO circuit where each $k$-local HUBO term requires $\sim 2(2k-3)$ two-qubit gates in the counterdiabatic layer. The inset shows the combinatorial upper-bound distribution of monomial orders for four instances $(N,M) = \{(12,4), (38,6), (61,8), (100,14)\}$ in the HUBO formulation. As problem size increases, $K'$ grows, extending the maximum interaction order $r_{\max}=2K'$ and shifting weight to higher-degree terms.
}
\label{fig:cvrp_2q_gate_scaling_beyond156}
\end{figure}

To assess the scalability of our approach beyond the limits of currently available quantum hardware, we focus on the CVRP use case as a representative large-scale industrial optimization problem. 
CVRP provides a natural benchmark for such an analysis due to its well-understood structure and its relevance in real-world logistics applications. Building on the resource estimates discussed above, we analyze how the circuit requirements scale for problem instances exceeding the qubit counts available on present-day devices. As illustrated in Fig.~\ref{fig:cvrp_2q_gate_scaling_beyond156}, the HUBO formulation offers a significant reduction in qubit count compared to QUBO through compact binary encoding, but this comes at the cost of increased two-qubit gates. This is a direct consequence of the increase in higher-order interaction terms for the HUBO formulation. This highlights the central trade-off: while HUBO enables access to larger problem sizes within a fixed qubit budget, the associated gate overhead becomes a critical bottleneck for scalability on current and near-term quantum hardware.

\begin{figure}
    \centering
    \includegraphics[width=\linewidth]{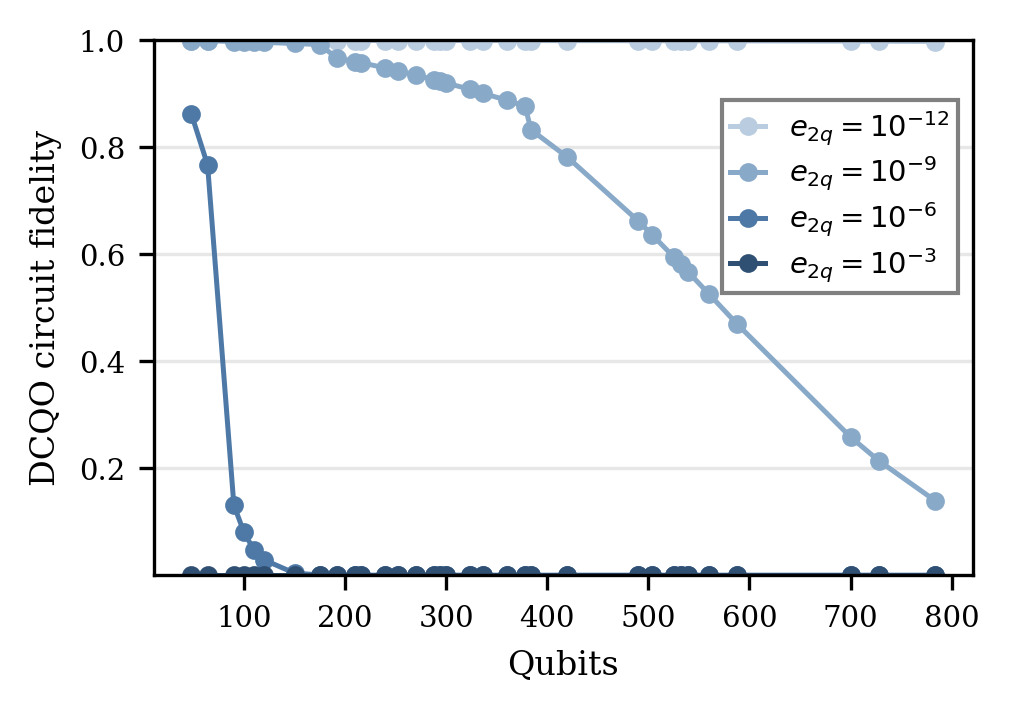}
    \caption{Circuit fidelity of HUBO-based DCQO circuits versus qubit number for different two-qubit error rates $e_{2q}$. The fidelity decays rapidly with CVRP problem size due to the increasing number of entangling gates, indicating the need for very low error rates to maintain reasonable circuit fidelity.}
    \label{fig:circuit_fidelity}
\end{figure}

To further quantify the impact of this trade-off, we analyze how circuit fidelity degrades with increasing problem size for HUBO-based implementations, as shown in Fig.~\ref{fig:circuit_fidelity}. 
As the problem size grows, the number of required entangling gates increases substantially due to the decomposition of higher-order terms, leading to an exponential suppression of the overall success probability~\eqref{eq:dcqo_fid} unless extremely low two-qubit error rates are achieved. 
In practice, this implies that even moderate-sized HUBO instances would require error rates approaching the logical-gate regime to maintain usable fidelity in the BF-DCQO algorithm. These results indicate that HUBO-based approaches are primarily relevant for future fault-tolerant hardware, while near-term implementations require problem decomposition techniques to remain tractable.

\begin{figure}
    \centering
\includegraphics[width=\linewidth]{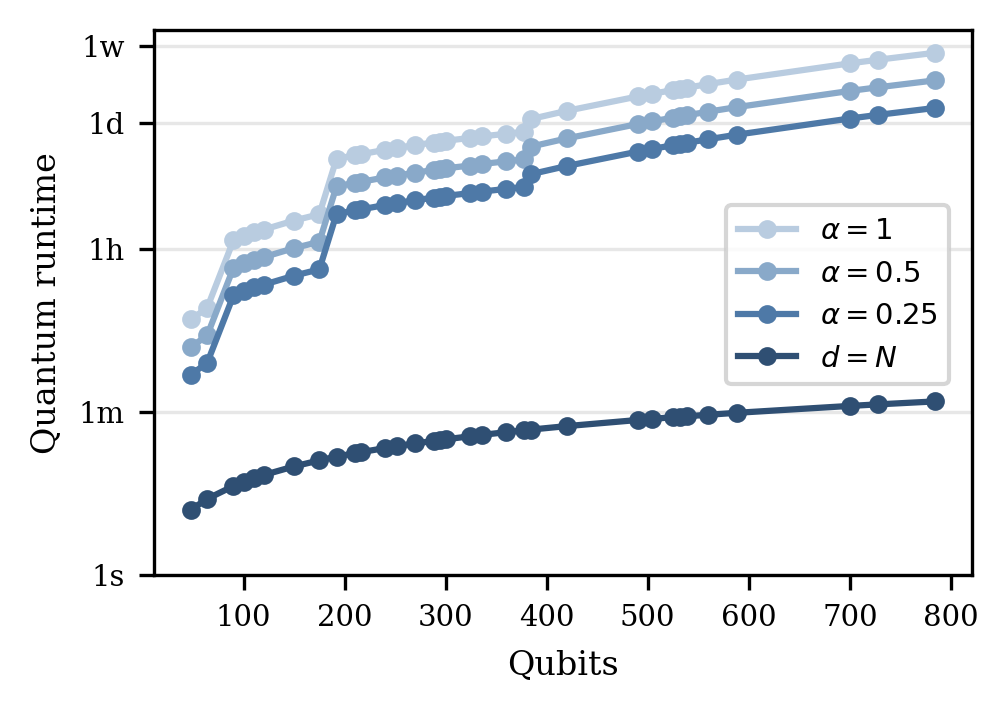}
    \caption{
    Estimated BF-DCQO quantum runtime versus qubit number for different depth reduction factors $\alpha$. The curve $d = N$ denotes the linear-depth lower bound. Runtime is computed assuming a fixed shot budget, $n_{\mathrm{iter}}=2$, two-qubit gate time $t_{2q}=50~\mu$s, and reset time $t_{\mathrm{reset}}=200~\mu$s, representative of an early fault-tolerant regime.}
    \label{fig:runtimes}
\end{figure}

To further assess practical feasibility in the fault-tolerant regime, we estimate the runtime of the quantum subroutine within a sequential hybrid workflow, rather than assuming a fully end-to-end quantum optimization. While fault-tolerant hardware mitigates the fidelity limitations discussed above, it introduces significant overhead in logical gate execution. In Fig.~\ref{fig:runtimes}, we consider a representative early fault-tolerant setting and assume a fixed shot budget on the order of $10^3$, corresponding to few-percent accuracy in expectation value estimation. Under these assumptions, the quantum runtime is dominated by circuit depth, and even moderate depth reductions (controlled by $\alpha$) lead to substantial improvements, whereas full-depth circuits ($\alpha=1$) quickly become impractical. In this sequential hybrid setting, the classical processing cost is typically comparable to or smaller than the quantum execution time and does not qualitatively change the overall scaling. Importantly, moving to slower hardware modalities, such as trapped-ion systems, or to deeper fault-tolerant regimes with larger code distances increases logical two-qubit gate times by one to two orders of magnitude, further amplifying the runtime of the quantum subroutine. These results indicate that practical implementations require bounded-depth or hybrid strategies rather than fully quantum end-to-end execution, even in future fault-tolerant settings.

\section{Scheduling problems}
\label{sec:scheduling}

Scheduling problems are a fundamental class of combinatorial optimization tasks that arise across industrial operations, where the challenge lies in assigning resources over time while satisfying complex and often competing constraints.
In automotive manufacturing, a representative example of such scheduling problems arises in the final stages of the production line, where the task is to assign different configuration criteria (e.g., sunroof, leather seats) to the incoming sequence of cars.
Since each criterion requires specific assembly stations with limited capacity, the order in which cars enter the assembly line has a direct impact on feasibility and throughput. In addition, the plant imposes a production target for each operational window, meaning a fixed number of cars must be processed within each scheduling horizon. These cars are temporarily buffered into storage lanes before entering the assembly line, and operational rules require that cars belonging to a given window be distributed as evenly as possible among all available storage units. Therefore, the sequencing task is not merely a permutation problem but a constrained optimization challenge, where the objective is to determine an ordering that minimizes the violation of operational rules while respecting machine capacity and storage distribution rules.

\subsection{Problem Formulation} 
To model the production scheduling problem as an optimization problem, we first distinguish between hard and soft constraints. Hard constraints enforce that each slot in the sequence can hold exactly one car and that each car is assigned to at most one slot, ensuring basic scheduling validity. Soft constraints arise from operational rules defined over configuration criteria and storage behavior. These include gap constraints, which specify a minimum separation between two cars with the same criteria (e.g., a minimum gap of 3 for the "sunroof" criterion requires that two sunroof-equipped cars appear at least three slots apart), as well as group constraints, which limit how frequently a given criterion may appear within a sliding window. The storage lane assignment introduces a secondary layer of constraints, with lower priority, that aims to evenly distribute cars across lanes to avoid local congestion.
Here, we address a set of industrial scheduling instances across multiple scales, grouped into four size categories (L, M, S and XS), with three instances per category to assess robustness and scalability; see Table~\ref{tab:instance_resources} for representative instance parameters and resource requirements across different size categories.
\begin{table*}
\centering
\caption{
Instance parameters and resource estimates for the scheduling problem across different size categories. For each size, the reported values correspond to a representative instance (maximum setting within the category). We report the number of jobs ($N$), slots ($M$), and storage lanes, along with the corresponding variable counts and maximum interaction order for the HUBO formulation. The QUBO representation is obtained via quadratization of the HUBO, introducing auxiliary variables and leading to an increased number of binary variables. Estimated two-qubit gate counts are computed assuming a single-step DCQO circuit, where each $k$-local term requires approximately $2(2k - 3)$ two-qubit gates after decomposition. The final column reports the number of variables in the corresponding ILP formulation.
}
\label{tab:instance_resources}
\vspace{0.5em}
\renewcommand{\arraystretch}{1.2}
\setlength{\tabcolsep}{5pt}
\begin{tabular}{c ccc ccc cc c}
\toprule
\multirow{2}{*}{\textbf{Instance}} 
& \multicolumn{3}{c}{\textbf{Parameters}} 
& \multicolumn{3}{c}{\textbf{HUBO}} 
& \multicolumn{2}{c}{\textbf{QUBO}} 
& \multicolumn{1}{c}{\textbf{ILP}} \\
\cmidrule(lr){2-4}\cmidrule(lr){5-7}\cmidrule(lr){8-9}\cmidrule(lr){10-10}
& $\bm{M}$ & $\bm{N}$ & \textbf{\# lanes}
& \textbf{\# vars} & \textbf{Max order} & \textbf{Two-qubit gates (est.)}
& \textbf{\# vars} & \textbf{Two-qubit gates (est.)}
& \textbf{\# vars} \\
\midrule
XS & 5 & 10 & 2 & 90 & 4 & 1834 & 94 & 1818 & 154 \\
S & 10 & 20 & 2 & 300 & 6 & 15046 & 329 & 14762 & 469 \\
M & 30 & 35 & 3 & 1380 & 21 & 149776 & 1523 & 147748 & 1959 \\
L & 30 & 100 & 5 & 3390 & 21 & 990732 & 3540 & 988740 & 4029 \\
\bottomrule
\end{tabular}
\end{table*}

We formulate the car sequencing task as a HUBO problem, where the objective is to minimize the total number of violations of the soft operational rules, each weighted by a penalty factor~$\lambda$ as specified by scheduling priorities, while strictly enforcing feasibility through the hard constraints.
The primary binary decision variables $y_{j\ell} \in \{0,1\}$ denote whether job $j \in \mathcal{J}$ is assigned to slot $\ell \in \mathcal{L}$, forming a one-hot encoding of the schedule. To capture higher-order penalties arising from gap and grouping rules, we introduce a few additional binary variables which indicate the activation of criterion $c$ at slot $\ell$, to reduce higher-order interactions directly in $y_{j\ell}$ variables.

The total HUBO cost function is then:
\begin{align}
\label{eq:hubo_scheduling}
H(\mathbf{y},\mathbf{x}) = & 
\lambda_{\mathrm{slot}} \sum_{\ell \in \mathcal{L}} 
\left(1 - \sum_{j \in \mathcal{J}} y_{j\ell}\right)^{2}
+ \lambda_{\mathrm{job}} \sum_{j \in \mathcal{J}} \sum_{\substack{\ell,\ell' \in \mathcal{L} \\ \ell<\ell'}} y_{j\ell}\,y_{j\ell'} \nonumber\\
& +\lambda_{\mathrm{link}} \sum_{c\in \mathcal{C}_\mathrm{gap} \cup\mathcal{C}_\mathrm{grp}\cup\mathcal{C}_\mathrm{lane}}\sum_{\ell \in \mathcal{L}} 
\left(x_{c\ell} -\sum_{j \in \mathcal{J}} m_{j,c} \, y_{j\ell} \right)^2 \nonumber \\
& + \sum_{c \in \mathcal{C}_{\mathrm{gap}}} H_{\mathrm{gap},c}(\mathbf{x}) + \sum_{c \in \mathcal{C}_{\mathrm{grp}}} H_{\mathrm{grp},c}(\mathbf{x}) \nonumber \\ 
& + \sum_{c \in \mathcal{C}_{\mathrm{grp+gap}}} H_{\mathrm{grp+gap},c}(\mathbf{x})
+ H_{\mathrm{FIFO}}(\mathbf{y}) \nonumber \\
& + \lambda_{\mathrm{lane}}
\sum_{\ell \in \mathcal{L}}
\left(
1-\sum_{c \in \mathcal{C}_{\mathrm{lane}}} x_{c\ell} \right)^{2} \nonumber\\
& + \sum_{c \in \mathcal{C}_{\mathrm{lane}}}\left( H_{\mathrm{gap},c}(\mathbf{x}) + H_{\mathrm{grp},c}(\mathbf{x}) \right),
\end{align}
where the operation term ``FIFO" stands for first-in-first-out, and the individual cost functions are elaborated in Appendix~\ref{app:scheduling}.
In accordance with operational requirements, the penalty weights for the hard scheduling constraints ($\lambda_{\mathrm{slot}}$, $\lambda_{\mathrm{job}}$, $\lambda_{\mathrm{link}}$ and $\lambda_{\mathrm{lane}}$) are set to be significantly higher than those associated with soft process rules, ensuring that feasibility is always prioritized over optimization of soft criteria.
Although most terms in the cost function~\eqref{eq:hubo_scheduling} are quadratic, as highlighted by the minor increase in the number of variables from HUBO to QUBO formulation~(Table~\ref{tab:instance_resources}), higher-order terms arise naturally from the group rule, where the order value is one more than the given group size.

\subsection{Rolling-horizon-based hybrid algorithm}
Given the limited qubit count and gate fidelities available on current quantum processors, solving large-scale industrial HUBO problems directly remains infeasible. Moreover, in the car sequencing use case, the cost function, i.e.\ a weighted number of rule violations, is predominantly local in nature, as most rules depend only on neighboring or nearby slots. To exploit this locality while accommodating present hardware constraints, we implement a hybrid rolling-horizon or block-decomposition algorithm. In this approach, all to-be-scheduled~(free) jobs are first coarsely sequenced based on certain heuristics. A trivial approach would be to sort according to their input sequence number~$e_{\mathrm{seq}}$ as it would also satisfy the FIFO rules, which penalize sequences with both non-increasing and rapidly increasing $e_{\mathrm{seq}}$.
The solver then iteratively selects the next subset of candidates~$N_{\mathrm{sub}}$ and constructs a reduced HUBO instance restricted to a local horizon of~$M_{\mathrm{sub}}$. This smaller binary optimization problem is then solved using an SA or BF-DCQO solver to determine a feasible local ordering. The scheduled~$M_{\mathrm{sub}}$ jobs are subsequently fixed and marked as completed, and the procedure repeats over the remaining horizon until the full sequence of length~$M$ is filled or no active jobs remain. The final global sequence is then constructed by concatenating these optimized local windows.

\subsection{Results}

We validate the proposed HUBO formulations using classical baselines prior to any comparison. Specifically, we employ a memetic Tabu search (MTS) algorithm~\cite{SILVA20211066}, a hybrid metaheuristic that combines local Tabu search with population-based exploration, and operates directly on the HUBO formulation without requiring quadratization. Due to the large size and complexity of the scheduling instances, BF-DCQO is not executed, even in simulation. The MTS algorithm yields a distribution of bitstrings representing the various low-cost assignments of the binary variables $y_{j\ell}$ and $x_{c\ell}$. Among these, we first filter out the infeasible bitstrings that violate hard scheduling constraints, and from the remaining feasible set, we select the bitstring corresponding to the lowest energy as the final solution to the combinatorial optimization problem. We then benchmark these results against an integer linear programming (ILP) formulation solved with Google OR-Tools. Finally, for all instances, we also employ the hybrid rolling-horizon workflow, using SA as the local optimizer for each subproblem. Despite the reduced qubit requirements of HUBO in this hybrid setting, the corresponding BF-DCQO circuit depths remain prohibitively large for classical simulation.

To quantitatively evaluate each obtained sequence, we compute the total number of rule violations across all criteria, including gap, group, lane, and sequence-related constraints. For the gap and group constraints, the violation metric is binary, i.e., each rule is considered either satisfied or violated, without accounting for the degree of deviation. In contrast, for the sequence (FIFO) constraints, we incorporate violation severity: whenever the assigned sequence numbers decrease within the car order, the magnitude of the difference between consecutive sequence numbers is taken as the violation measure. Additionally, excessive increases beyond a predefined threshold are also penalized.

\begin{figure*}
    \centering
    \includegraphics[width=\linewidth]{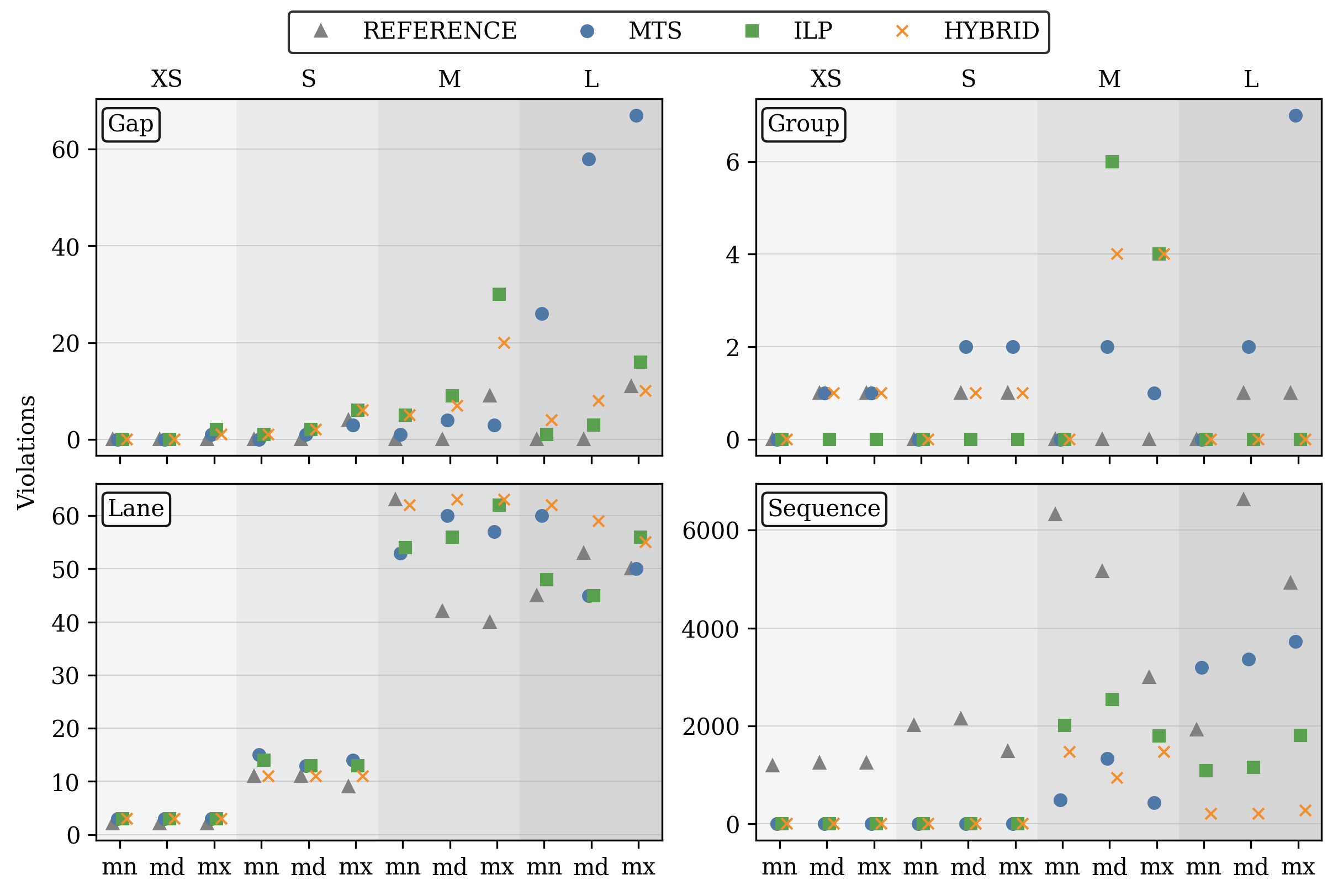}
    \caption{
Comparison of constraint violations across all scheduling instances for different solution methods: baseline reference (REFERENCE), memetic Tabu search (MTS), integer linear programming (ILP), and the hybrid rolling-horizon approach (HYBRID). Results are grouped by instance size (XS, S, M, L), with three instances per category (minimum, median, and maximum). Each panel corresponds to a specific constraint type: gap, group, lane, and sequence (FIFO). The plots illustrate how violation counts vary across methods and problem scales, enabling a direct comparison of solution quality and robustness.
}  \label{fig:scheduling_results}
\end{figure*}

As shown in Fig.~\ref{fig:scheduling_results}, the number of constraint violations increases with instance size, with sequence (FIFO) violations exhibiting the most significant growth. In particular, the reference solution achieves low violations for local constraints such as group rules, but at the cost of very large FIFO violations, highlighting the trade-off between local feasibility and global sequence consistency. The MTS solver effectively reduces FIFO violations across all instances; however, its performance degrades for larger problem sizes, which can be attributed to the increasing polynomial order of the HUBO formulation (up to order 21), making the optimization landscape more challenging. In contrast, the ILP solver consistently enforces local constraints such as gap and group rules but accumulates larger sequence violations, especially at scale. Lane-related violations also increase with instance size and remain comparatively challenging across all methods, with no single solver clearly dominating. The hybrid rolling-horizon approach, which employs simulated annealing (SA) as a local optimizer, achieves a more balanced performance by effectively reducing both global and local violations. Notably, for medium-sized (M) instances, where the problem remains within a tractable regime, MTS is able to better handle higher-order group constraints compared to both ILP and the hybrid method.

\section{Conclusion}
\label{sec:conclusion}

We assess the potential of quantum computing through three representative combinatorial optimization case studies motivated by real-world scenarios in logistics and manufacturing industries.
For each use case, we develop higher-order binary optimization (HUBO) formulations and systematically compare their scalability and near-term implementability against established quadratic (QUBO) models. Across the transportation-routing cases (QUEST and CVRP), higher-order terms emerge from compact encoding of problem variables, enabling a substantially reduced variable count and, consequently, qubit footprint relative to QUBO formulations. In the assembly-line scheduling case, by contrast, higher-order structure originates from accurately representing highly correlated real-world constraints that are otherwise ``softened" by quadratic approximations. 
We validate these formulations using classical solvers across instances of increasing size, and additionally benchmark small routing instances using classically simulated BF-DCQO.

Our resource analysis emphasizes that qubit count is only one dimension of practical feasibility. Successful digital quantum optimization also depends critically on circuit depth and two-qubit gate counts, which are further influenced by hardware connectivity. In practice, implementability is therefore governed not only by the number and layout of available qubits (quantity), but also by the fidelity (quality) and coherence times (lifetime) of entangling gate operations. We extend this feasibility analysis using the CVRP use case as a representative large-scale benchmark, given its relevance in real-world logistics applications. For a realistically scaled instance with 100 customers, we estimate that a faithful implementation of digitized counterdiabatic (DCQO) circuits would require average two-qubit gate error rates below $10^{-10}$. Such error levels are only realistic in the logical-qubit regime enabled by quantum error correction, rather than on present-day physical devices.
The increased duration of logical gate operations further amplifies runtime costs, rendering fully end-to-end quantum optimization impractical at scale and motivating a sequential hybrid approach that combines quantum and classical routines.

Looking forward, the practical impact of quantum optimization in industrial settings will depend on aligning problem structure with hardware capabilities. Many automotive optimization tasks, such as assembly-line sequencing with gap and group rules or routing problems with capacity and ordering constraints, are inherently higher-order and therefore naturally suited to HUBO formulations. For such problems, quantum computing provides a natural framework to represent and process higher-order interactions without requiring reduction to quadratic form.

In the near term, given the limited qubit counts and gate fidelities of available devices, problem-decomposition-based hybrid approaches will be essential to map large industrial instances onto tractable subproblems. Even in the early fault-tolerant regime, the increased duration of logical gate operations can lead to prohibitively large runtimes for end-to-end quantum optimization. This motivates a sequential hybrid approach, where quantum routines are used to refine solutions obtained from classical optimization, rather than replacing them entirely. As quantum hardware matures, these hybrid strategies may provide a practical pathway to exploit the favorable qubit scaling of HUBO formulations, while avoiding the expensive runtimes of end-to-end quantum optimization, thereby enabling representations beyond the constraints imposed by quadratic encoding.

\section*{Acknowledgements}
The authors thank their collaborators at Volkswagen AG for suggesting use cases and valuable discussions. We also acknowledge useful discussions within Kipu Quantum.

\appendix
\section{HUBO formulation of QUEST use case}\label{app:quest_HUBO}

Analogously to the TSP formulation in Ref.~\cite{Glos2020SpaceEfficient}, for each breaker $b\in\{1,\dots,B\}$, we represent the assigned surfer by a $K$-bit binary string
\[
x_b=(x_{b,0},\dots,x_{b,K-1}),
\qquad x_{b,k}\in\{0,1\},
\]
where
\[
K=\lceil \log_2 S\rceil
\]
is the number of bits required to encode the $S$ available surfers. For each surfer label $s\in\{0,\dots,S-1\}$, we define $\beta_s\in\{0,1\}^K$ as the fixed binary representation of surfer $s$.

To determine whether breaker $b$ is assigned surfer $s$, we introduce the selector function
\begin{equation}
\delta(b,s)=\prod_{k=0}^{K-1}\bigl(1-(x_{b,k}-\beta_s^k)^2\bigr),
\end{equation}
which evaluates to $1$ when the binary string $x_b$ matches the encoding of surfer $s$, and to $0$ otherwise.

The optimization cost is represented by a matrix $\omega\in\mathbb{R}^{S\times B}$, where each entry $\omega_{s,b}$ denotes the cost associated with assigning surfer $s$ to breaker $b$. This cost includes the aerodynamic benefit as well as penalties for mismatches in arrival times and preferred speeds, as described in \cite{Onah2025QUEST}. The objective is to determine a bijective assignment between breakers and surfers that minimizes the total matching cost.

The total QUEST cost is modeled as
\begin{equation}
H_{\mathrm{obj}}
=
\sum_{b=1}^{B}\sum_{s=0}^{S-1}\omega_{s,b}\,\delta(b,s),
\end{equation}
so that a cost contribution is added only when breaker $b$ is assigned surfer $s$.

To ensure feasible assignments, we introduce the following penalty terms. First, binary encoding may generate labels that do not correspond to any physical surfer whenever $S$ is not a power of two. Such invalid assignments are penalized by
\begin{equation}
H_{\mathrm{valid}}
=
\sum_{b=1}^{B}
\left(
1-\sum_{s=0}^{S-1}\delta(b,s)
\right)^2,
\end{equation}
which enforces that each breaker is assigned a valid surfer label.

Second, to enforce that no two breakers are assigned the same surfer, we define the equality indicator
\begin{equation}
H_{\mathrm{eq}}(x,x')
=
\prod_{k=0}^{K-1}\left(1-(x_k-x'_k)^2\right),
\end{equation}
which evaluates to $1$ iff the two encoded surfer assignments are identical. The corresponding uniqueness penalty is
\begin{equation}
H_{\mathrm{unique}}
=
\sum_{b=1}^{B-1}\sum_{b'=b+1}^{B}
H_{\mathrm{eq}}(x_b,x_{b'}).
\end{equation}

The complete QUEST Hamiltonian is therefore
\begin{align}
H(\mathbf{x})
&=
H_{\mathrm{obj}}(\mathbf{x})
+\lambda_{\mathrm{valid}} H_{\mathrm{valid}}(\mathbf{x})
+\lambda_{\mathrm{unique}} H_{\mathrm{unique}}(\mathbf{x}),
\end{align}
where $\mathbf{x}=\{x_{b,k}\}$ denotes the full collection of binary variables, and $\lambda_{\mathrm{valid}},\lambda_{\mathrm{unique}}$ are penalty coefficients chosen large enough that invalid or non-unique assignments are energetically suppressed.

This selector-based formulation makes the QUEST encoding structurally parallel to the CVRP formulation discussed below: in both cases, compact binary encodings are used for discrete assignments, selector functions activate cost terms associated with physical labels, and additional penalties enforce feasibility of the encoded solution.

Note: For clarity, the three use cases employ separate notation conventions. In QUEST, breakers and surfers are indexed by $b$ and $s$, respectively. In CVRP, vehicles, route positions, and nodes are indexed by $v$, $t$, and $i$. In the scheduling problem, jobs, slots, and criteria are indexed by $j$, $\ell$, and $c$. Binary-encoded decision variables are denoted by $x$ in QUEST and CVRP, whereas in scheduling the primary assignment variables are denoted by $y_{j\ell}$ and the auxiliary criterion indicators by $x_{c\ell}$.

\section{HUBO formulation of CVRP use case}\label{app:CVRP_HUBO}

For each node $i \in \{0,\dots,n-1\}$, we define $\beta_i \in \{0,1\}^{K'}$ as the fixed binary representation of node $i$. To determine whether the binary variables at a given position represent a particular node, we introduce a selector function defined as
\begin{equation} \label{eq:selector}
\delta(v,t,i)=\prod_{k=0}^{K'-1}\bigl(1-(x_{v,t,k}-\beta _i ^ k)^2\bigr),
\end{equation}
which evaluates to $1$ when the binary values $(x_{v,t,0}, \dots, x_{v,t,K'-1})$ correspond exactly to the binary code of node $i$, and to $0$ otherwise.

The travel cost between nodes is represented by the nonnegative matrix $W \in \mathbb{R}^{n \times n}$, where each element $W_{ij}$ corresponds to the distance or cost of traveling from node $i$ to node $j$. Each customer $i$ has a demand $q_i \ge 0$, while the depot has $q_0 = 0$. For simplicity, all vehicles are assumed to share the same capacity $Q$. The objective is to determine the set of routes that minimizes total travel cost while satisfying customer, capacity, and route continuity constraints.

The corresponding travel-cost objective is:
\begin{equation}
\label{eq:obj}
H_{\mathrm{obj}} =
\sum_{v=0}^{M-1}\sum_{t=0}^{L-2}\sum_{i=0}^{n-1}\sum_{j=0}^{n-1}
W_{ij}\,\delta(v,t,i)\,\delta(v,t+1,j),
\end{equation}
where the product of selector terms ensures that a cost is added only when vehicle $v$ travels from node $i$ at position $t$ to node $j$ at position $t+1$. In the CVRP and QUEST formulations, the highest-order terms arise from contributions that act on two encoded labels simultaneously, i.e., on two $K'$-bit substrings of the full bitstring. Since each selector function depends on $K'$ binary variables, products of two selectors can involve up to $2K'$ binary variables. Therefore, the maximum interaction order in the expanded HUBO is $2K'$.

To ensure feasible routing solutions, several penalty terms are introduced. Firstly, each customer must be visited exactly once across all vehicles and route positions:
\begin{equation} \label{eq:visit}
H_{\mathrm{visit}}=\sum_{i=1}^{n-1}\Bigl(\sum_{v=0}^{M-1}\sum_{t=0}^{L-1}\delta(v,t,i)-1\Bigr)^2,
\end{equation}

Each vehicle must also perform only one continuous trip, meaning that once it returns to the depot, it cannot leave again:
\begin{equation} \label{eq:mono}
H_{\mathrm{mono}}=\sum_{v=0}^{M-1}\sum_{t=0}^{L-2}\delta(v,t,0)\bigl(1-\delta(v,t+1,0)\bigr),
\end{equation}

Because binary encoding can represent node indices that exceed the number of actual nodes, additional invalid combinations must be penalized to ensure only physical nodes are assigned:
\begin{equation} \label{eq:valid}
H_{\mathrm{valid}}=\sum_{v=0}^{M-1}\sum_{t=0}^{L-1}\sum_{i=n}^{2^{K'}-1}\delta(v,t,i),
\end{equation}

Finally, to ensure that the total demand assigned to each vehicle does not exceed its capacity, a quadratic penalty term is introduced:
\begin{equation} \label{eq:cap}
H_{\mathrm{cap}}
=
\sum_{v=0}^{M-1}
\Bigl(\sum_{t=0}^{L-1}\sum_{i=1}^{n-1} q_i\,\delta(v,t,i)-Q\Bigr)^2.
\end{equation}
This term softly enforces the capacity constraint, although exact inequalities can be implemented by introducing slack variables if required.

\section{HUBO formulation of scheduling use case}
\label{app:scheduling}
The given use case is a typical assembly-line scheduling problem where all jobs/items follow the same route of machines. Given items $\mathcal{J}=\{1,\dots,N\}$ and slots $\mathcal{L}=\{1,\dots,M\}$, we define a binary variable 
\begin{equation}
    y_{j\ell} =
\begin{cases}
1, & \text{if job $j$ is assigned to slot $\ell$,} \\
0, & \text{otherwise.}
\end{cases}
\end{equation}
Similarly, for a given set of criteria $\mathcal{C}=\{1,\dots,F\}$ and membership $m_{j,c}\in\{0,1\}$ indicating whether job $j$ has the criterion $c$, we define criterion-at-slot indicators as 
\begin{equation}
    x_{c\ell} =
\begin{cases}
1, & \text{if criterion $c$ is assigned to slot $\ell$,} \\
0, & \text{otherwise.}
\end{cases}
\end{equation}
For gap and group rules, we opt to treat the corresponding $x_{c\ell}$ as independent variables with additional constraints linking them to $y_{j\ell}$.
The HUBO cost function is defined over all $y_{j\ell}$ and $x_{c\ell}$ variables, with penalty terms arising from both hard, i.e.\ feasibility, constraints and soft constraints, i.e.\ rule violations. 

The hard constraints, with high penalty coefficients, ensure that a sequence of items is operationally feasible. For instance, an item can be scheduled at most once. Similarly, at most one item can be assigned in each slot of the sequence. Conversely, we can add a strong penalty corresponding to exactly one item per slot, and take care of the under-assignment option in the feasibility check. The penalty terms are defined as follows.
\begin{enumerate}
    \item Exactly one job per slot:
Each slot $\ell$ must contain exactly one job, with
\begin{equation}
 H_{\mathrm{slot}}(y) = 
\lambda_{\mathrm{slot}} \sum_{\ell \in \mathcal{L}} 
\left(1 - \sum_{j \in \mathcal{J}} y_{j\ell}\right)^{2}.   
\end{equation}

\item  At most one slot per job:
Each job $j$ can be assigned to at most one slot $\ell$, with
\begin{equation}
H_{\mathrm{job}}(y) =
\lambda_{\mathrm{job}} 
\sum_{j \in \mathcal{J}}
\sum_{\substack{\ell,\ell' \in \mathcal{L}\\ \ell < \ell'}} 
y_{j\ell}\,y_{j\ell'}.
\end{equation}

\item For each criterion~$c$ requiring additional free variables, the job assigned at slot~$\ell$ should also carry that criterion, which is enforced via a  quadratic penalty
\begin{align}
\label{eq:job_crit_link}
H_{\mathrm{link}}(x,y)
=
\lambda_{\mathrm{link}}
\sum_{c\in\mathcal{C}}
\sum_{\ell\in\mathcal{L}}
\left(
x_{c\ell}
-
\sum_{j\in\mathcal{J}} m_{j,c}\,y_{j\ell}
\right)^2 ,
\end{align}
ensuring consistency between job assignments and activated criteria.

\end{enumerate}
The penalty coefficients for these hard constraints are chosen according to operational priorities, ensuring $\lambda_{\mathrm{slot}} \approx \lambda_{\mathrm{job}} \approx \lambda_{\mathrm{link}} \gg \lambda_{\mathrm{soft}}$, to guide optimization without violating feasibility.

We introduce soft sequencing constraints comprising gap rules enforcing spacing between criteria,  group rules controlling contiguous block lengths, and  constraints based on input sequence numbers; for some criteria, gap and group rules apply simultaneously. 
The penalty terms are as follows.
\begin{enumerate}
\item Gap constraints.
For a criterion $c$ requiring either ``exact'' or ``at least'' spacing $d_c$
(in slots) between two jobs holding the same criterion, we define
\begin{equation}
H^{\text{exact}}_{\mathrm{gap}}(x)
=
\lambda_{\mathrm{gap}}
\sum_{\ell=1}^{M}
\left(
\sum_{h=1}^{d_c} x_{c,\ell}\,x_{c,\ell-h}
\;+\;
x_{c,\ell}\big(1 - x_{c,\ell-d_c-1}\big)
\right),
\end{equation}
\begin{equation}
\label{eq:gap_atleast}
H^{\text{atleast}}_{\mathrm{gap}}(x)
=
\lambda_{\mathrm{gap}}
\sum_{\ell=1}^{M-d_c}
\sum_{h=1}^{d_c}
x_{c,\ell}\,x_{c,\ell+h}.
\end{equation}
In the implementation, the slot indexing is extended to allow indices $\ell \le 0$, which represent positions in the known production history prior to the optimization horizon. Variables $x_{c,t}$ with $t \le 0$ are treated as fixed constants determined by the historical sequence.

\item Group constraints.
For a criterion \(c\), group rules may enforce either an ``exact" or an ``at most" contiguous block.
An exact group length \(U_c\) is enforced by penalizing all maximal runs shorter or longer than \(U_c\) as
\begin{align}
H^{\mathrm{exact}}_{\mathrm{grp}}(x) 
=&
\lambda_{\mathrm{under}}\sum_{r=1}^{U_c-1}
\sum_{\ell=1}^{M-r}
(1-x_{c,\ell-1})
%\left(
\prod_{h=0}^{r-1} x_{c,\ell+h}
%\right)
(1-x_{c,\ell+r}) \nonumber\\
&+
\lambda_{\mathrm{over}}
\sum_{\ell=1}^{M-U_c}
\prod_{h=0}^{U_c} x_{c,\ell+h}.
\end{align}
Alternatively, an at-most group rule forbids runs of more than \(U_c\) consecutive activations as
\begin{align}
H_{\mathrm{grp}}^{\mathrm{atmost}}(x)
=
\lambda_{\mathrm{grp}}
\sum_{\ell=1}^{M-U_c}
\prod_{h=0}^{U_c} x_{c,\ell+h}.
\end{align}
Similar to the previous case, variables  $x_{c,\ell}$ with $\ell \le 0$ are treated as fixed constants determined by the known history.
These group constraints introduce higher-order interaction terms in the Hamiltonian, with the order scaling with the group size.

\item Simultaneous group and gap constraints.
For a criterion~\(c\) with an exact gap \(d_c(\neq0)\) between consecutive groups of length exactly \(U_c\), we define
\begin{widetext}
\begin{align}
H_{\mathrm{grp+gap}}^{\mathrm{exact}}(x)
&=
H_{\mathrm{grp}}^{\mathrm{exact}}(x)
\;+\;
\lambda_{\mathrm{short}}
\sum_{\ell=1}^{M-U_c}
\left(
\prod_{h=0}^{U_c-1} x_{c,\ell+h}
\right)
\left(
\sum_{i=0}^{d_c-1} x_{c,\ell+U_c+i}
\right)
% \\
% &\quad
+\;
\lambda_{\mathrm{long}}
\sum_{\ell=1}^{M-(U_c+d_c+1)}
\left(
\prod_{h=0}^{U_c-1} x_{c,\ell+h}
\right)
\left(
\prod_{i=0}^{d_c} \bigl(1-x_{c,\ell+U_c+i}\bigr)
\right)
x_{c,\ell+U_c+d_c+1}.
\end{align}
\end{widetext}

Another combination that appears is an exact gap of zero and a non-zero value for at most group. In this case, the gap constraint can be safely ignored as it is redundant.
Also, when \(U_c = 1\) and an ``at least'' gap \(d_c\) is imposed, the group constraint is redundant, since the gap condition already enforces isolated occurrences separated by at least \(d_c\) slots.

\item Sequence-number (FIFO) constraints.
In addition to criterion-based rules, we impose soft constraints based on the input sequence number \(e_j\) of item \(j\). Ideally, $e_i$ should be lower than $e_k$ if job~$i$ is placed before job~$k$, and the gap between the sequence numbers should not exceed a certain value.
\begin{equation}
    H_{\mathrm{FIFO}}^{\mathrm{down}}(y)=
\lambda_{\mathrm{down}}
\sum_{\ell=1}^{M-1}
\sum_{i\in\mathcal{J}}
\sum_{k\in\mathcal{J}}
\max\{0,\,e_i-e_k\}\; y_{i\ell}\,y_{k,\ell+1},
\end{equation}
penalizes local FIFO violations in proportion to the backward jump \(e_i-e_k\).
\(H_{\mathrm{FIFO}}^{\mathrm{step}}\) penalizes adjacent ``step-ups'' exceeding a threshold \(g\), while \(H_{\mathrm{FIFO}}^{\mathrm{span}}\) penalizes schedules whose end-to-end sequence-number span exceeds \(D\). These are expressed as
\begin{align}
H_{\mathrm{FIFO}}^{\mathrm{step}}(y)
&=
\lambda_{\mathrm{step}}
\sum_{\ell=1}^{M-1}
\sum_{i\in\mathcal{J}}
\sum_{k\in\mathcal{J}}
\mathbb{I}\!\left[e_k-e_i>g\right]\; y_{i\ell}\,y_{k,\ell+1},\\[1mm]
H_{\mathrm{FIFO}}^{\mathrm{span}}(y)
&=
\lambda_{\mathrm{span}}
\sum_{i\in\mathcal{J}}
\sum_{k\in\mathcal{J}}
\mathbb{I}\!\left[e_k-e_i>D\right]\; y_{i1}\,y_{kM},
\end{align}
respectively,
where $\mathbb{I}[\cdot]$ denotes the indicator function.
These terms collectively capture FIFO behavior by penalizing both local inversions and large deviations from the desired sequence ordering.
\end{enumerate}

Lane assignments are modeled as additional criteria attached to items. Each job may be associated with one or more admissible lanes, and the selected lane at a given slot is represented through the corresponding lane criterion variable. As in the general criterion construction, we enforce hard linking constraints to ensure consistency between the job assigned to a slot and the activated lane criterion~\eqref{eq:job_crit_link}. 
Additionally, we enforce that exactly one lane is selected at each slot through the quadratic penalty
\begin{equation}
H_{\mathrm{lane}}(x)
=
\lambda_{\mathrm{lane}}
\sum_{\ell \in \mathcal{L}}
\left(
1 - \sum_{c \in \mathcal{C}_{\mathrm{lane}}} x_{c\ell}
\right)^2.
\end{equation}
In practice, lane constraints can be viewed as a special case of criterion-based constraints, where the combination of group and gap rules simplifies to a pure gap constraint. Therefore, lane-related soft penalties are implemented using only the corresponding gap terms~\eqref{eq:gap_atleast}.

\section{Two-qubit gate estimation}\label{app:polynomial_scaling}
This appendix details the estimation procedure underlying the resource analysis presented in \S\ref{sec:routing}.
To quantify the quantum resources required by the proposed HUBO Hamiltonians, we estimate the number of two-qubit gates needed for their implementation. 
Since the underlying decision variables are represented in binary form, the Hamiltonians acquire a polynomial structure in the binary variables, which naturally gives rise to higher-order interaction terms. The corresponding gate cost is thus determined by the largest interaction order appearing in the expansion and by the distribution of monomial contributions across different orders.

Each decision variable taking one of $n$ discrete values is encoded in binary form using
\begin{equation}
K' = \lceil \log_2 n \rceil.
\end{equation}
The selector function defined in Eq.~\ref{eq:selector} is therefore a polynomial of degree at most $K'$ in the corresponding binary variables. Consequently, products of selector functions generate higher-order monomials in the expanded Hamiltonian. In the CVRP and QUEST formulations, the highest-order contributions arise from terms coupling two binary-encoded assignments, such as products of selector functions or equality-indicator terms, so the maximum interaction order is
\begin{equation}
r_{\max} = 2K'.
\end{equation}

After expansion, the Hamiltonian can be written as a sum of monomials of different degrees, where each degree-$r$ monomial corresponds to an $r$-body interaction term. Let $N_r$ denote the number of degree-$r$ monomial contributions. Each $r$-body $Z\cdots Z$ term in the problem Hamiltonian gives rise to $\mathcal{O}(r)$ counterdiabatic terms with a single $Y$ insertion (e.g., $Y Z \cdots Z$), each of which naively requires $2(r-1)$ two-qubit gates when implemented via standard CNOT ladders~\cite{decomp2023}. By exploiting shared entangling structure and canceling intermediate operations across these terms, the total cost can be reduced to $2(2r-3)$ two-qubit gates per $r$-body interaction. Assuming this cost per $r$-body term, the total number of two-qubit gates is estimated as
\begin{equation}
N_{2q} = \sum_{r=2}^{r_{\max}} N_r \, 2(2r - 3).
\label{eq:D3}
\end{equation}

The quantities $N_r$ are estimated using a combinatorial upper-bound argument based on the number of binary variables participating in each local Hamiltonian contribution. In particular, if a local term depends on at most $m$ binary variables, then the number of degree-$r$ monomials it can generate is bounded by
\begin{equation}
N_r \leq \binom{m}{r}.
\end{equation}
In the CVRP case, local terms involving products of two selector functions depend on at most $2K'$ binary variables, so one obtains the bound $N_r \leq \binom{2K'}{r}$. More generally, the final order distribution is obtained by applying this counting argument to each class of Hamiltonian terms and summing the resulting contributions. These estimates represent upper bounds, as cancellations and repeated monomials may reduce the actual gate count.

\bibliography{ref}

\end{document}